\documentclass[doublecol,12pt]{epl2}
\usepackage{amsmath,colonequals}
\usepackage{amssymb}
\usepackage{graphicx}
\usepackage{epsfig}
\usepackage{bm}
\usepackage{color}

\newfont{\sss}{cmmi10 at 20.74pt}

\def\spose#1{\hbox to 0pt{#1\hss}}
\def\simlt{\mathrel{\spose{\lower 3pt\hbox{$\mathchar"218$}}
     \raise 2.0pt\hbox{$\mathchar"13C$}}}
\def\simgt{\mathrel{\spose{\lower 3pt\hbox{$\mathchar"218$}}
     \raise 2.0pt\hbox{$\mathchar"13E$}}}
\def\0{\mbox{\boldmath$\displaystyle\mathbf{0}$}}

\def\be{\begin{equation}}
\def\ee{\end{equation}}
\def\bea{\begin{eqnarray}}
\def\eea{\end{eqnarray}}

\newcommand{\units}[1]{\ensuremath{\,\mathrm{#1}}}
\newcommand{\ce}{\colonequals}

\title{Neutrino oscillations with disentanglement of a neutrino from its partners}
\author{D. V. Ahluwalia\thanks{E-mail: \email{dharamvir.ahluwalia@canterbury.ac.nz}} \and S.  P. Horvath\thanks{E-mail: \email{sebastian.horvath@pg.canterbury.ac.nz}}}

\shortauthor{D. V. Ahluwalia \etal}
\institute{Department of Physics and Astronomy,
Rutherford Building, University of Canterbury,
Private Bag 4800, Christchurch 8140, New Zealand}


\abstract{We bring attention to the fact that in order to understand existing data on neutrino oscillations, and to design future experiments, it is imperative to appreciate the role of quantum entanglement. Once this is accounted for, the resulting energy-momentum conserving phenomenology requires a single new parameter related to disentanglement of a neutrino from its partners. This parameter may not be CP symmetric. We illustrate the new ideas, with potentially measurable effects, in the context of a novel experiment recently proposed by Gavrin, Gorbachev, Veretenkin, and Cleveland. The strongest impact of our ideas is on the resolution of various anomalies in neutrino oscillations and on neutrino propagation in astrophysical environments.}

\begin{document}
\maketitle
\section{Introduction}
%%change-SPH -- some minor re-wording.  
%The experimental confirmation of neutrino oscillations has not only resolved the problem of solar neutrino anomaly~\cite{Ahmad:2002jz} but it also has  implications for a whole range of phenomena, such as the evolution and explosions of type II supernovae~\cite{Ahluwalia:1997dv}. It has also been realised that neutrino oscillations provide a set of flavour oscillation clocks and they redshift precisely as required by the general-relativistic theory of gravity (GR)~\cite{Ahluwalia:1996ev,Ahluwalia:1998jx,Konno:1998kq}. In precision experiments the insights gained from these studies can test the interface of GR and quantum realms in a manner that complements gravitationally induced effects in neutron~\cite{PhysRevLett.34.1472,PhysRevA.56.1767} and atom interferometry~\cite{Chu:1999}.
The experimental confirmation of neutrino oscillations has not only resolved the problem of the solar neutrino anomaly~\cite{Ahmad:2002jz}, shed light on atmospheric neutrino data~\cite{Ashie:2004mr}, inspired various reactor and accelerator based experiments,
but it also has  implications for a whole range of phenomena, such as the evolution and explosions of type II supernovae~\cite{Ahluwalia:1997dv,Raffelt:2003en,Choubey:2006aq,Woosley:2006ie,Hidaka:2007se,Marek:2007gr,Fuller:2009zz,Dighe:2009nr,Duan:2009cd,Chakraborty:2009ej}. Indeed, implications as far reaching as the interface between general relativity (GR) and the quantum realm have become accessible by the discovery that neutrino oscillations provide a set of flavour oscillation clocks that redshift precisely as required by GR~\cite{Ahluwalia:1996ev,Ahluwalia:1998jx,Konno:1998kq,Grossman:1996eh,Camacho:1999hv,Wudka:2000rf,Adak:2000tp,Pereira:2000kq,Linet:2002wp,Crocker:2003cw,Lambiase:2005gt}. These insights can, at least in principle, be tested by precision experiments that complement the experiments on gravitationally induced effects in neutron~\cite{PhysRevLett.34.1472,PhysRevA.56.1767} and atom interferometry~\cite{Chu:1999}.

Underlying all these successes and investigations is the Pontecorvo assumption that a neutrino produced in a weak interaction mediated process may be considered as a linear superposition of different mass eigenstates.
\begin{equation}
  \vert\nu_\ell \rangle = \sum_i U_{\ell i} \vert\nu_i\rangle\label{eq:pontecorvo}
\end{equation}
where the index $\ell \in\{e,\mu,\tau\}$ represents the three flavours, the index $ i\in \{1,2,3\}$ spans the three mass eigenstates, and the $3\times3$ unitary matrix $U$ defines and distinguishes the three flavours. The CP-conjugated states are identified with the antineutrino, and are given by
\begin{equation}
  \vert\bar\nu_\ell \rangle = \sum_i U^\ast_{\ell i} \vert\bar\nu_i\rangle\label{eq:pontecorvo-anti}
\end{equation}
Indeed such a phenomenological ansatz is able to provide a good broad brush understanding for much of the existing data. 

However, theoretically, such a highly successful ansatz cannot be the complete description of the kinematic structure of neutrino oscillations; a quiet reflection reveals that it violates conservation of the energy-momentum four vector. This is something that is not allowed for weak interaction mediated processes which are known to be invariant under spacetime translations. This fact was first realised by Goldman in 1996~\cite{Goldman:1996yq}, and a while later by Nauenberg~\cite{Nauenberg:1998vy}. Apart from some notable exceptions, the physics community seems to have neglected this remarkable observation. A small set of other papers devoted to the subject of energy-momentum conservation in neutrino oscillations appear to have created more heat than light. However, the publication of a recent paper by Cohen, Glashow, and Ligeti has now made it unambiguously clear that the conventional understanding of neutrino oscillations is not entirely tenable and that the suggested revision must be incorporated in any analysis of neutrino oscillations~\cite{Cohen:2008qb}. This revision makes a quantum entanglement of neutrinos with the accompanying decay partners a necessary and unavoidable element of the physics of neutrino oscillations. This has fundamental consequences for the correct interpretation of the relevant data.

The focus in Ref.~\cite{Cohen:2008qb} was entirely on Raghavan's proposed experiment~\cite{Raghavan:2008cs} and on the variations in decay times observed in the GSI experiment~\cite{Litvinov:2008rk}. Here we show that implications of the new insights extend far beyond their original scope. Towards this end we first review the energy-momentum conserving neutrino oscillation formalism in a concrete setting. We then argue how it enters and revises our understanding of neutrino oscillations and its existing puzzles. Once this is appreciated, a whole range of new questions arise. Answers to these questions may affect not only the design and interpretations of terrestrial experiments but it may also lead to a better understanding of the neutrino-induced~\textemdash~with quantum entanglement a necessary ingredient~\textemdash~astrophysical processes. We briefly discuss these aspects.

\section{Quantum entanglement and neutrino oscillations}

We are inspired by the following observation.
Whenever a mass eigenstate decays into a set of other mass eigenstates the relevant conservation laws may induce a quantum entanglement. For instance, in the EPR-like decay of a spin zero particle at rest into two spin one half particles (in the singlet state), the conservation of angular momentum forces the spin projections of the decay products to be entangled. Such an entanglement is quite robust~\cite{PhysRevA.76.052110} and it is destroyed, e.g., when the entangled attribute is subjected to a measurement or an interaction.

\subsection{Neutrino identification puzzle and neutrino oscillations}
The just mentioned EPR entanglement is induced by the invariance of the underlying interactions under rotations. For neutrinos, the operative interaction is the weak interaction of the standard model. It is invariant under space, as well as time, translations. Therefore, with appropriate caveats and parenthetic remarks about the width of wave functions, etc., a neutrino produced in the decay of a mass eigenstate (say a pion)~\textemdash~which we assume is an eigenstate of both energy and momentum~\textemdash~must also be in an eigenstate of energy and momentum. However, since each of the $\vert\nu_i\rangle$ in Eq.~(\ref{eq:pontecorvo}) has different mass, $\vert\nu_\ell\rangle$ cannot simultaneously be an eigenstate of energy and momentum.\footnote{Same remarks apply to $\vert\bar\nu_\ell\rangle$ and Eq.~(\ref{eq:pontecorvo-anti}).} This fact forbids $\vert\nu_\ell \rangle$ of Eq.~(\ref{eq:pontecorvo}) to be identified with a neutrino produced in a weak interaction mediated process~\cite[cf. discussion in Sec. 2]{Cohen:2008qb}.~We shall call this circumstance the neutrino identification puzzle. This concern was first emphasised by Cohen, Glashow,  and Ligeti~\cite{Cohen:2008qb}.

To be concrete, consider the production of $\nu_\mu$ and $\bar{\nu}_\mu$   in the following CP conjugated processes\footnote{We note the decay of $\mu^\pm$ explicitly for later reference below. For simplicity, we assume $\pi^\pm$ at rest.}
\begin{subequations}
  \begin{equation}
    \begin{picture}(-20,30)(100,-20)
      \put(0,0){$\pi^+$}
      \put(15,2.5){\vector(1,0){15}} 
      \put(35,0){$\nu_\mu + \mu^+$}
      \put(60,-7.5){\line(0,-1){10}}
      \put(60,-17.5){\vector(1,0){10}}
      \put(80,-20){$ e^{+} + \nu_\text{e} + \bar{\nu}_\mu $}
    \end{picture}\label{eq:piondecaya}
  \end{equation}
  and
  \begin{equation}
    \begin{picture}(-20,30)(100,-20) % 140,30 picture size in pt; -20,-20 moves center left 20 pt
      \put(0,0){$\pi^-$} %origin after moving
      \put(15,2.5){\vector(1,0){15}} % 15 start of arrow, 2.5 takes it 2.5 pt up \put{where}{what} (1,0)= direction horizontal right, {-1,0}=horizontal left,{1,1} = up etc. Last enter 15 is 15 pt length of the vector
      \put(35,0){$\bar{\nu}_\mu + \mu^-$}
      \put(60,-7.5){\line(0,-1){10}}
      \put(60,-17.5){\vector(1,0){10}}
      \put(80,-20){$ e^{-} + \bar{\nu}_\text{e} + {\nu}_\mu $}
    \end{picture}\label{eq:piondecayb}
  \end{equation}
\end{subequations}
The neutrino identification puzzle is the assertion that neutrinos and antineutrinos that are thus produced cannot be linear superpositions of different mass eigenstates (in the sense of equations~(\ref{eq:pontecorvo}) and (\ref{eq:pontecorvo-anti})). 

The energy-momentum four vector conserving correct identifications, as was first argued by Goldman~\cite{Goldman:1996yq}~\textemdash~and later emphasised by Nauenberg on the one hand~\cite{Nauenberg:1998vy}  and Cohen, Glashow,  and Ligeti on the other hand~\cite{Cohen:2008qb}~\textemdash~require quantum entanglement between the decay products. For the above-considered  $\pi^\pm$ decay at rest, the neutrino-muon entanglement becomes manifest by re-writing (in fact, by correcting) $\nu_\mu + \mu^+$ and $\bar{\nu}_\mu + \mu^-$ in equations~(\ref{eq:piondecaya}-\ref{eq:piondecayb}) as
\begin{subequations}
  \begin{equation}
   \sum_i U_{\mu i} \underbrace{\left\vert \nu_i\right\rangle}_{\left\vert\sqrt{\mathbf{k}_i^2+m_i^2},\;\mathbf{k}_i,\;m_i^2\right\rangle} \otimes \underbrace{\left\vert\mu^+\right\rangle}_{\left\vert E_i,\;-\mathbf{k}_i,\; m_\mu^2\right\rangle} \label{eq:entanglementa}
  \end{equation}
  and
  \begin{equation}
   \sum_i U^\ast_{\mu i} \underbrace{\left\vert \bar{\nu}_i\right\rangle}_{\left\vert \sqrt{\mathbf{k}_i^2+m_i^2},\;\mathbf{k}_i,\;m_i^2\right\rangle} \otimes \underbrace{\left\vert\mu^-\right\rangle}_{\left\vert E_i,\;-\mathbf{k}_i,\; m_\mu^2\right\rangle}\label{eq:entanglementb}
  \end{equation}
\end{subequations}
The notational details are adapted from Ref.~\cite{Goldman:1996yq}.\footnote{Here, $\mathbf{k}_i$ represents the momentum associated with $i$th mass eigenstate, $\vert\nu_i\rangle$, and $m_i$ is the mass of the $i$th mass eigenstate.} These identifications are consistent with the results of references~\cite{Nauenberg:1998vy,Cohen:2008qb}). The energy-momentum conserving generalisation to the case when $\pi^\pm$ are not at rest is straightforward. It can be found in Ref.~\cite{Cohen:2008qb}.

Thus, the resolution of the neutrino identification puzzle resides in the replacement of the Pontecorvo suggestion contained in equations~(\ref{eq:pontecorvo}-\ref{eq:pontecorvo-anti}) by the Goldman suggested source-dependent identifications such as the ones displayed above. It is worth noting that the original title of Ref.~\cite{Goldman:1996yq} was `Source Dependence of Neutrino Oscillations'. It was only in the published version, roughly a decade and half later, that it changed to `Neutrino Oscillations and Energy-Momentum Conservation'. The energy-momentum conserving definition of neutrinos, and their quantum entanglement, is necessarily source dependent. For instance, the $\nu_\mu$ produced in a $\pi^+$ decay has a different quantum entanglement than a $\nu_\mu$ created in the process $\mu^- \to e^- +\bar\nu_e+ \nu_\mu$ or those obtained from $e^- + e^+ \to \nu_\ell +\bar\nu_\ell$. 

For the discussion that follows we shall call states of neutrinos and antineutrinos represented by equations~(\ref{eq:pontecorvo}-\ref{eq:pontecorvo-anti}) as \emph{Pontecorvo states} while the states of the type represented by equations~(\ref{eq:entanglementa}-\ref{eq:entanglementb}) will be called \emph{Goldman states}. This allows the discussion so far to be summarised as: weak interactions produce Goldman states, not Pontecorvo states.\footnote{A parenthetic remark: once the Goldman suggestion is taken seriously one must in fact also incorporate the entanglement induced by the conservation of angular momentum (weak interactions are also invariant under rotations). This forces the states given in equations~(\ref{eq:entanglementa}-\ref{eq:entanglementb}) to be replaced by their proper singlet state expressions which contain not only the left-transforming neutrino with the negative helicity, but also the left-transforming neutrino with the positive helicity. The latter may mimic some of the signatures of a sterile neutrino (similar remarks hold true for antineutrinos). This is a subject that must be treated in a subsequent study. In order to not raise too many issues at once the remainder of this paper takes equations~(\ref{eq:entanglementa}-\ref{eq:entanglementb}) without any additional modifications.} This is a simple, and simultaneously profound, resolution of the neutrino identification puzzle. Further, Cohen, Glashow,  and Ligeti (CGL)~\cite{Cohen:2008qb}) show that flavour oscillations do not occur for the Goldman states. For flavour oscillations to occur a projection to the Pontecorvo states must be made. They arrive at the following conclusion (quoted below, verbatim)
\begin{quote}
\begin{enumerate}
  \item[\small cgl-1] When the decay products of an initial state of well-defined
    momentum evolve without further interaction no oscillation phenomena
    appear.
  \item[\small cgl-2] To realize oscillations the
    neutrino mass eigenstates must be disentangled.
\end{enumerate}
\end{quote}
Since these results are rather counterintuitive, an element of repetition is in order:  cgl-1 translates to the fact that weak interactions produce Goldman states, while cgl-2 is the statement that for oscillations to occur a projection to the Pontecorvo states must occur.

%\section{Modification to Neutrino Oscillation Pheno\-menology}

%Interaction of the Goldman states with the environment results in a projection to Pontecorvo states. 
This has the consequence that, for a Goldman state produced at the laboratory time $t=0$ and projected to a Pontecorvo state at time $t=\tau_d$, the standard neutrino oscillation probability is modified. For the simple and representative  $2\times 2$ setting, the modified $\nu_e\to \nu_e$ probability then reads (assuming ultra-relativistic neutrinos)
\begin{eqnarray}
  \hspace{-25pt}&&P^\prime(\nu_e \to\nu_e) = H(c \tau_d -r)\hfill\nonumber \\
  \hspace{-25pt}&& + H(r-c \tau_d)\left[1 -\sin^2(2\theta) \sin^2\left(1.27 \Delta m^2  \frac{\left[L - c \tau_d\right)}{E_\nu}\right)\right]
  \label{eq:mp}
\end{eqnarray}
where $H(t)$ is the standard Heaviside step function, $c= 3\times 10^8 \units{m/s}$, $r$ is the 
running position along the neutrino trajectory (measured in meters, $r=0$ is the creation region),
the source-detector distance $L$ is in meters, the mass-squared difference $\Delta m^2$ is in $\mbox{eV}^2$, and the neutrino energy $E_\nu$ is in $\mbox{MeV}^2$. For a given ensemble of neutrinos, $\tau_d$ may be a distribution function depending on the details of a given set up. For example, if one of the decay products is short lived, $\tau_d$ may depend on the lifetime of the entangled partner.

\section{Implications for neutrino oscillations in laboratory and astrophysical environments}

%%change-SPH -- some modifications to make clear that this is now in a different context to KARMEN-LSND
%Expression (\ref{eq:mp}) indicates that the KARMEN-LSND anomaly can be immediately resolved if $\tau_d$ for these experiments is close to the KARMEN source-detector distance~\cite{Eitel:1999gt} in natural units. 

%The modification to the neutrino oscillation probability can also be detected more directly in the proposed Gavrin, Gorbachev, Veretenkin, and Cleveland (GGVC) proposal~\cite{Gavrin:2010qj} if the disentanglement time $\tau_d$ is determined by the dimensions of the reentrant tube (and hence to the size of the $^{51}\mbox{Cr}$ source). Figure 3 of  Ref.~\cite{Gavrin:2010qj} gives the ratio $R_2/R_1$ for the $\nu_e$ capture rates for the inner and outer zones in the GGVC proposal. Those calculations assume un-modified version of eq.~(\ref{eq:mp}). In Figure 1 we provide the ratio 
%\begin{equation}
%  \beta\colonequals\frac{R_2^\prime/R_1^\prime}{R_2/R_1}
%\end{equation} 
%with $R_2^\prime/R_1^\prime$ calculated using the modified oscillation probability given in Eq.~(\ref{eq:mp}).  An immediate consequence that follows from expression (\ref{eq:mp}) is that the KARMEN-LSND anomaly can be easily resolved if $\tau_d$ for these experiments is close to the KARMEN source-detector distance in natural units~\cite{Eitel:1999gt}. 

An immediate consequence that follows from expression (\ref{eq:mp}) is that the KARMEN-LSND anomaly can be easily resolved if $\tau_d$ for these experiments is close to the KARMEN source-detector distance in natural units~\cite{Eitel:1999gt}.

Turning to the proposal by Gavrin, Gorbachev, Veretenkin, and Cleveland (GGVC)~\cite{Gavrin:2010qj}, one finds an experimental candidate for which the proposed modification to the neutrino oscillation probability would become directly detectable, provided the disentanglement time $\tau_d$ is determined by the dimensions of the reentrant tube (and hence to the size of the $^{51}\mbox{Cr}$ source). Figure 3 of Ref.~\cite{Gavrin:2010qj} gives the ratio $R_2/R_1$ for the $\nu_e$ capture rates for the inner and outer zones of the GGVC proposal. Those calculations assume the un-modified version of Eq.~(\ref{eq:mp}). Figure 1 below details the ratio 
\begin{equation}
\beta\colonequals\frac{R_2^\prime/R_1^\prime}{R_2/R_1}
\label{eq:beta}
\end{equation} 
with $R_2^\prime/R_1^\prime$ calculated using the modified oscillation probability given in Eq.~(\ref{eq:mp}). In order to gain insight into the dependence of $\beta$ on the disentanglement time, Figure 2 shows a re-evaluation of Eq.~(\ref{eq:beta}) with $\tau_d$ increased by an order of magnitude; for details, the reader is referred to the figure captions.

It is worth emphasising that the detectability of a quantum entanglement in this experiment presupposes that the MiniBooNE~\cite{AguilarArevalo:2010wv} and the LSND experiments~\cite{Athanassopoulos:1996jb,Athanassopoulos:1997pv}, the low capture rates in the Ga source experiments~\cite{Giunti:2010zu}, and antineutrino reactor anomaly~\cite{Mention:2011rk}, all suggest existence of an eV\textsuperscript{2} range mass-squared difference \emph{and} that the disentanglement time $\tau_d$ is roughly related to the source size (for `low-mass' source case\footnote{Clearly, we are guided here by calculations on the mean free path of neutrons and the known sizes of the critical mass in nuclear reactors/devices.}). It would, however, be extremely helpful if an \emph{ab initio} estimate of $\tau_d$ was made through the development of a theory of quantum disentanglement in nuclear media. 

%%change-SPH -- as I included a reference to figure 2 in the text above, I removed it from the caption of figure 1 below. 
\begin{figure}
  \includegraphics[width=240pt]{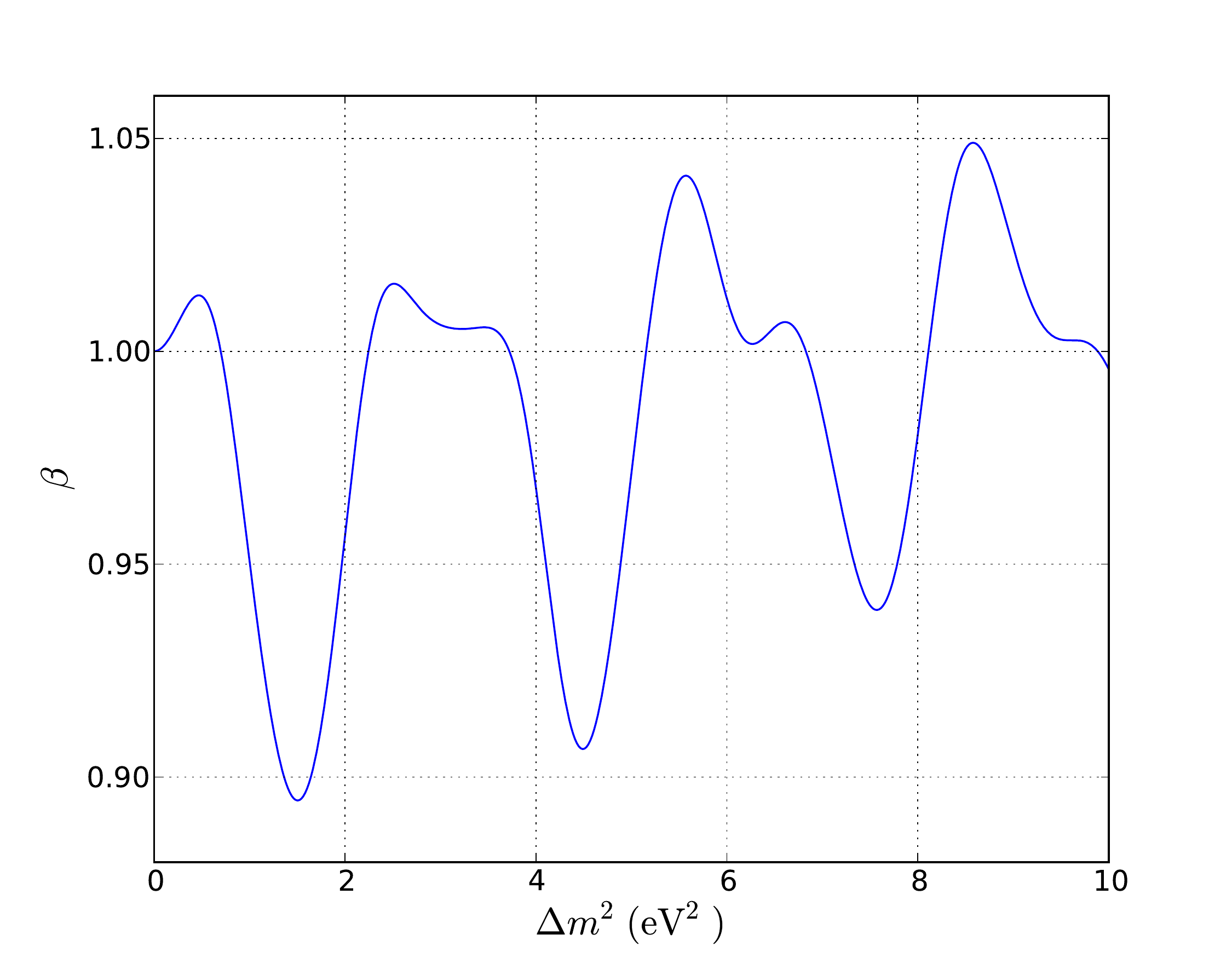}
  \caption{In drawing this figure we have chosen $\tau_d \ce \tau_0 =  0.105\units{meters} \times c$. This choice corresponds to the radial size of the $^{51}\mbox{Cr}$ source that is used in the GGVC proposal. In the event that the measured $\beta$ turns out to be extremely close to unity, one may turn the argument around and either measure, or place limits on, $\tau_d$. To match the GGVC proposal, the plot corresponds to the case $\sin^2 2\theta = 0.3$.} 
\end{figure}

%\section{Interpretational consequences}
\begin{figure}
  \includegraphics[width=240pt]{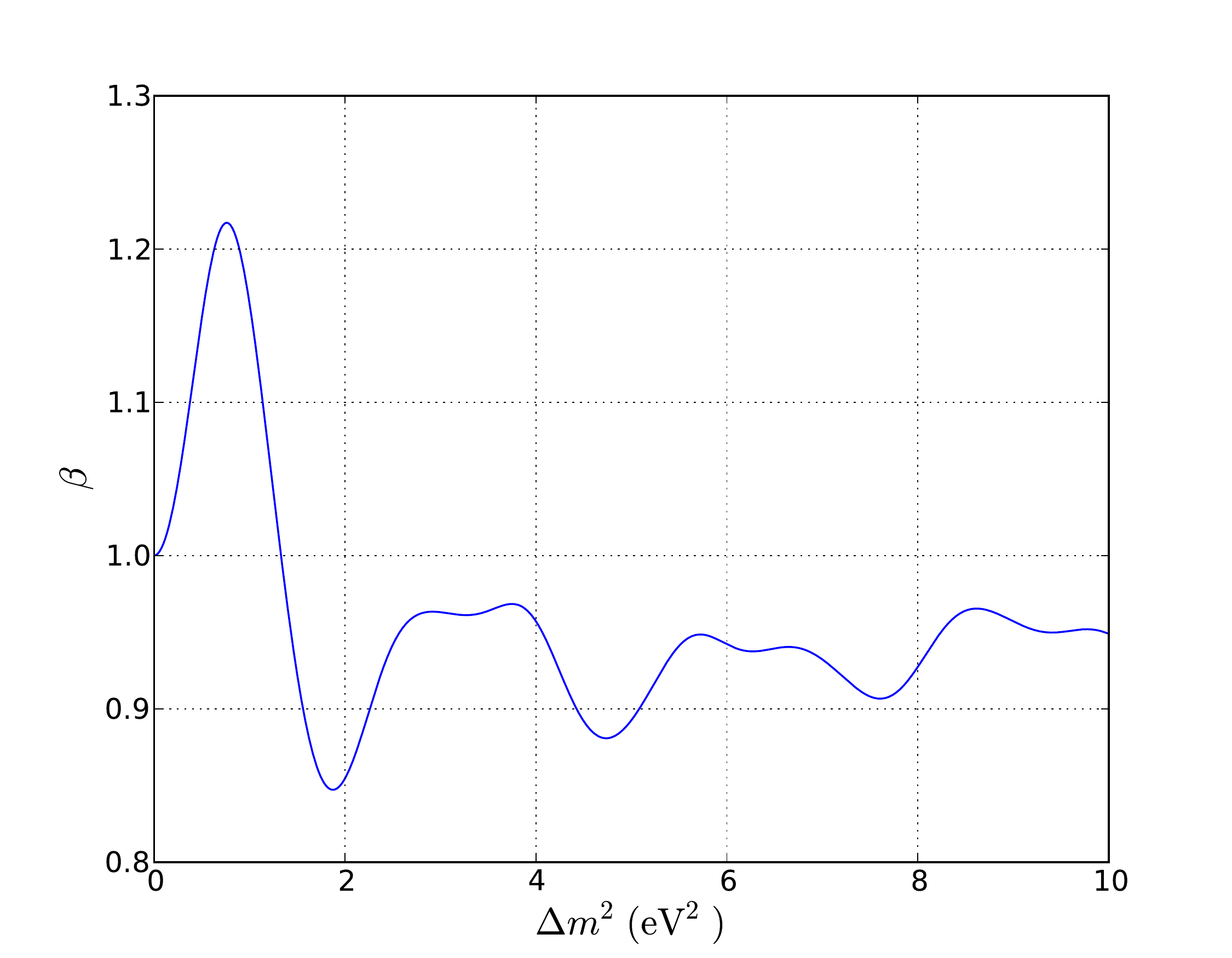}
  \caption{A complementary plot to Figure 1 for a disentanglement time of $\tau_d = 10\times \tau_0$. To match the GGVC proposal, the plot corresponds to the case $\sin^2 2\theta = 0.3$.
  In our calculations we have taken care of the fact that $90\%$ of the decays of the $^{51}\mbox{Cr}$ source produce  $750 \units{keV}$ neutrinos, while $10\%$ of the decays create $430\units{keV}$ neutrinos. The advantage of presenting results in terms of $\beta$, rather than in terms of $R_2^\prime/R_1^\prime$ versus  $R_2/R_1$, lies in the fact that in calculation for $\beta$ we do not need detailed information about the detector efficiency, it cancels out.
} 
\end{figure}

%%change-SPH -- I rewrote the first sentence or two in the direct context of MiniBooNE as a general neutrino experiment may not have a beam stop etc. 
%Furthermore, it is not \emph{a priori} obvious that $\tau_d$ is CP symmetric. To see this, let us return to the decay chain in equations (\ref{eq:piondecaya}) and (\ref{eq:piondecayb}). If the disentanglement  time $\tau_d > \tau_\mu$, the muon lifetime, then $\tau_d$ is necessarily CP asymmetric. This is so because the $e^+$ produced in $\mu^+$ decay,  annihilate with electrons in the beam stop, whereas $e^-$ produced in $\mu^-$ decay, do not. These asymmetric interactions of the entangled partners of neutrinos and antineutrinos produced in decays of $\pi^\pm$ with the beam stop effect the projection of the Goldman states to Pontecorvo states via a built-in asymmetry in $\tau_d$. 
Furthermore, it is not \emph{a priori} obvious for many experiments that $\tau_d$ is CP symmetric. As an example, consider the neutrino source of the MiniBooNE experiment which is produced by the decay chain in equations (\ref{eq:piondecaya}) and (\ref{eq:piondecayb}). If the disentanglement time $\tau_d > \tau_\mu$, the muon lifetime, then $\tau_d$ is necessarily CP asymmetric. This is because the $e^+$ produced in the $\mu^+$ decay, annihilate with electrons in the beam stop, whereas the $e^-$ produced in the $\mu^-$ decay, do not. These asymmetric interactions of the entangled partners of neutrinos and antineutrinos, produced in decays of $\pi^\pm$, with the beam stop affect the projection of the Goldman states to Pontecorvo states via a built-in asymmetry in $\tau_d$. 

Interestingly, the MiniBooNE source-detector distance is $1.8\units{\mu s}$, in natural units. This is quite close to the muon lifetime $\tau_\mu \approx 2.20\units{\mu s}$. This observation when combined with the possibility that the CP-asymmetry of $\tau_d$ for the MiniBooNE setting may not be same as for the LSND setting (where the source-detector distance is about $0.1\units{\mu s}$), could turn out to be an important ingredient in reconciling these two experiments. 

%%change-SPH -- move this section just before the conclusion, this way the statement of affecting all neutrino oscillation experiments comes after all the explicit examples. Additionally, some moderately minor rewriting/modification of a few sentences. 
%It is now apparent that quantum entanglement has the potential to affect interpretation of all neutrino oscillations experiments. Additionally, since neutrino oscillation play pivotal role in type II supernovae explosions, and in nucleosynthesis of light elements~\cite{Yoshida:2006qz}, the effects such as those implicit in the modified neutrino oscillation phenomenology may have significant impact for those studies. For the pion-decay neutrinos the disentanglement time $\tau_d$, interestingly, depends on the electromagnetic interactions, and not on weak interactions, and is therefore expected to be ``small''. On the other hand, for the neutrinos produced in the $e^\pm$ annihilations in neutron stars $\tau_d$, may be quite `large' as the mean free path of these neutrinos in neutron-star media is of the order of a kilometer. In this latter instance, there are no charged entangled partners that are electromagnetically charged. This fact may dramatically alter the evolution of neutron stars and may require recalculation of neutrino-oscillations effects in the explosion of type II supernovae.
It is thus apparent that quantum entanglement has the potential to affect the interpretation of all neutrino oscillations experiments. Additionally, since neutrino oscillations play a pivotal role in type II supernovae explosions, and in the nucleosynthesis of light elements~\cite{Yoshida:2006qz}, effects such as those implicit in the modified neutrino oscillation phenomenology may have a significant impact for those studies. For pion-decay neutrinos the disentanglement time $\tau_d$, interestingly, depends on the electromagnetic interactions, and not on the weak interactions, and is therefore expected to be ``small''. On the other hand, for neutrinos produced in the $e^\pm$ annihilations in neutron stars, $\tau_d$ may be quite ``large'' as the mean free path of these neutrinos in neutron-star media is of the order of a kilometer. In this latter instance, there are no entangled partners that are electromagnetically charged. This fact may dramatically alter the evolution of neutron stars and may require a recalculation of neutrino-oscillation effects in the explosion of type II supernovae.

\section{Conclusion}

We have provided a brief review of why simultaneous conservation of energy and momentum requires the decay partners of neutrinos to be entangled. No flavour oscillations occur until this entanglement is destroyed. This has consequences that have the potential to resolve the KARMEN-LSND conflict, change expectations of the GGVC proposal, introduce non-intrinsic CP violation, and affect neutrino propagation and oscillations in astrophysical environments.

%\bibliographystyle{JHEP} % Please do not remove the titles. They contain important information.
%\bibliography{Westport-Spring2010}

\begin{thebibliography}{10}

\bibitem{Ahmad:2002jz}
{\bf SNO Collaboration} Collaboration, Q.~Ahmad {\em et.~al.}, {\it {Direct
  evidence for neutrino flavor transformation from neutral current interactions
  in the Sudbury Neutrino Observatory}},  {\em Phys.Rev.Lett.} {\bf 89} (2002)
  011301, [\href{http://xxx.lanl.gov/abs/nucl-ex/0204008}{{\tt
  nucl-ex/0204008}}].

\bibitem{Ashie:2004mr}
{\bf Super-Kamiokande Collaboration} Collaboration, Y.~Ashie {\em et.~al.},
  {\it {Evidence for an oscillatory signature in atmospheric neutrino
  oscillation}},  {\em Phys.Rev.Lett.} {\bf 93} (2004) 101801,
  [\href{http://xxx.lanl.gov/abs/hep-ex/0404034}{{\tt hep-ex/0404034}}].

\bibitem{Ahluwalia:1997dv}
D.~V. Ahluwalia, {\it {Addendum to: Gen. Rel. Grav. 28 (1996) 1161, First Prize
  Essay for 1996: Neutrino Oscillations and Supernovae}},  {\em Gen. Rel.
  Grav.} {\bf 36} (2004) 2183--2187,
  [\href{http://xxx.lanl.gov/abs/astro-ph/0404055v1}{{\tt
  astro-ph/0404055v1}}].

\bibitem{Raffelt:2003en}
G.~G. Raffelt, M.~T. Keil, R.~Buras, H.-T. Janka, and M.~Rampp, {\it {Supernova
  neutrinos: Flavor-dependent fluxes and spectra}},
  \href{http://xxx.lanl.gov/abs/astro-ph/0303226}{{\tt astro-ph/0303226}}.

\bibitem{Choubey:2006aq}
S.~Choubey, N.~P. Harries, and G.~G. Ross, {\it {Probing neutrino oscillations
  from supernovae shock waves via the IceCube detector}},  {\em Phys. Rev.}
  {\bf D74} (2006) 053010, [\href{http://xxx.lanl.gov/abs/hep-ph/0605255}{{\tt
  hep-ph/0605255}}].

\bibitem{Woosley:2006ie}
S.~Woosley and T.~Janka, {\it {The physics of core-collapse supernovae}},  {\em
  Nature Physics} (2006) [\href{http://xxx.lanl.gov/abs/astro-ph/0601261}{{\tt
  astro-ph/0601261}}].

\bibitem{Hidaka:2007se}
J.~Hidaka and G.~M. Fuller, {\it {Sterile Neutrino-Enhanced Supernova
  Explosions}},  {\em Phys.Rev.} {\bf D76} (2007) 083516,
  [\href{http://xxx.lanl.gov/abs/0706.3886}{{\tt arXiv:0706.3886}}].

\bibitem{Marek:2007gr}
A.~Marek and H.-T. Janka, {\it {Delayed neutrino-driven supernova explosions
  aided by the standing accretion-shock instability}},  {\em Astrophys.J.} {\bf
  694} (2009) 664--696, [\href{http://xxx.lanl.gov/abs/0708.3372}{{\tt
  arXiv:0708.3372}}]. * Brief entry *.

\bibitem{Fuller:2009zz}
G.~M. Fuller, A.~Kusenko, and K.~Petraki, {\it {Eosphoric sterile neutrinos,
  supernovae, and the galactic positrons}},  {\em Phys. Lett.} {\bf B670}
  (2009) 281--284, [\href{http://xxx.lanl.gov/abs/0806.4273}{{\tt
  arXiv:0806.4273}}].

\bibitem{Dighe:2009nr}
A.~Dighe, {\it {Supernova neutrino oscillations: What do we understand?}},
  {\em J.Phys.Conf.Ser.} {\bf 203} (2010) 012015,
  [\href{http://xxx.lanl.gov/abs/0912.4167}{{\tt arXiv:0912.4167}}].

\bibitem{Duan:2009cd}
H.~Duan and J.~P. Kneller, {\it {Neutrino flavour transformation in
  supernovae}},  {\em J. Phys.} {\bf G36} (2009) 113201,
  [\href{http://xxx.lanl.gov/abs/0904.0974}{{\tt arXiv:0904.0974}}].

\bibitem{Chakraborty:2009ej}
S.~Chakraborty, S.~Choubey, S.~Goswami, and K.~Kar, {\it {Collective Flavor
  Oscillations Of Supernova Neutrinos and r-Process Nucleosynthesis}},  {\em
  JCAP} {\bf 1006} (2010) 007, [\href{http://xxx.lanl.gov/abs/0911.1218}{{\tt
  arXiv:0911.1218}}].

\bibitem{Ahluwalia:1996ev}
D.~V. Ahluwalia and C.~Burgard, {\it {Gravitationally Induced Quantum
  Mechanical Phases and Neutrino Oscillations in Astrophysical Environments}},
  {\em Gen. Rel. Grav.} {\bf 28} (1996) 1161--1170,
  [\href{http://xxx.lanl.gov/abs/gr-qc/9603008}{{\tt gr-qc/9603008}}].

\bibitem{Ahluwalia:1998jx}
D.~V. Ahluwalia and C.~Burgard, {\it {Interplay of gravitation and linear
  superposition of different mass eigenstates}},  {\em Phys. Rev.} {\bf D57}
  (1998) 4724--4727, [\href{http://xxx.lanl.gov/abs/gr-qc/9803013}{{\tt
  gr-qc/9803013}}].

\bibitem{Konno:1998kq}
K.~Konno and M.~Kasai, {\it {General relativistic effects of gravity in quantum
  mechanics: A case of ultra-relativistic, spin 1/2 particles}},  {\em Prog.
  Theor. Phys.} {\bf 100} (1998) 1145--1157,
  [\href{http://xxx.lanl.gov/abs/gr-qc/0603035}{{\tt gr-qc/0603035}}].

\bibitem{Grossman:1996eh}
Y.~Grossman and H.~J. Lipkin, {\it {Flavor oscillations from a spatially
  localized source: A simple general treatment}},  {\em Phys. Rev.} {\bf D55}
  (1997) 2760--2767, [\href{http://xxx.lanl.gov/abs/hep-ph/9607201}{{\tt
  hep-ph/9607201}}].

\bibitem{Camacho:1999hv}
A.~Camacho, {\it {Flavor-oscillation clocks, continuous quantum measurements
  and a violation of Einstein equivalence principle}},  {\em Mod. Phys. Lett.}
  {\bf A14} (1999) 2545--2556,
  [\href{http://xxx.lanl.gov/abs/gr-qc/9911112}{{\tt gr-qc/9911112}}].

\bibitem{Wudka:2000rf}
J.~Wudka, {\it {Mass dependence of the gravitationally-induced wave-function
  phase}},  {\em Phys. Rev.} {\bf D64} (2001) 065009,
  [\href{http://xxx.lanl.gov/abs/gr-qc/0010077}{{\tt gr-qc/0010077}}].

\bibitem{Adak:2000tp}
M.~Adak, T.~Dereli, and L.~H. Ryder, {\it {Neutrino oscillations induced by
  space-time torsion}},  {\em Class. Quant. Grav.} {\bf 18} (2001) 1503--1512,
  [\href{http://xxx.lanl.gov/abs/gr-qc/0103046}{{\tt gr-qc/0103046}}].

\bibitem{Pereira:2000kq}
J.~G. Pereira and C.~M. Zhang, {\it {Some remarks on the neutrino oscillation
  phase in a gravitational field}},  {\em Gen. Rel. Grav.} {\bf 32} (2000)
  1633--1637, [\href{http://xxx.lanl.gov/abs/gr-qc/0002066}{{\tt
  gr-qc/0002066}}].

\bibitem{Linet:2002wp}
B.~Linet and P.~Teyssandier, {\it {Quantum phase shift and neutrino
  oscillations in a stationary, weak gravitational field}},
  \href{http://xxx.lanl.gov/abs/gr-qc/0206056}{{\tt gr-qc/0206056}}.

\bibitem{Crocker:2003cw}
R.~M. Crocker, C.~Giunti, and D.~J. Mortlock, {\it {Neutrino interferometry in
  curved spacetime}},  {\em Phys. Rev.} {\bf D69} (2004) 063008,
  [\href{http://xxx.lanl.gov/abs/hep-ph/0308168}{{\tt hep-ph/0308168}}].

\bibitem{Lambiase:2005gt}
G.~Lambiase, G.~Papini, R.~Punzi, and G.~Scarpetta, {\it {Neutrino optics and
  oscillations in gravitational fields}},  {\em Phys. Rev.} {\bf D71} (2005)
  073011, [\href{http://xxx.lanl.gov/abs/gr-qc/0503027}{{\tt gr-qc/0503027}}].

\bibitem{PhysRevLett.34.1472}
R.~Colella, A.~W. Overhauser, and S.~A. Werner, {\it Observation of
  gravitationally induced quantum interference},  {\em Phys. Rev. Lett.} {\bf
  34} (Jun, 1975) 1472--1474.

\bibitem{PhysRevA.56.1767}
K.~C. Littrell, B.~E. Allman, and S.~A. Werner, {\it Two-wavelength-difference
  measurement of gravitationally induced quantum interference phases},  {\em
  Phys. Rev. A} {\bf 56} (Sep, 1997) 1767--1780.

\bibitem{Chu:1999}
A.~Peters, K.~Y. Chung, and Chu, {\it Measurement of gravitational acceleration
  by dropping atoms},  {\em Nature} {\bf 400} (1999) 849--851.

\bibitem{Goldman:1996yq}
T.~Goldman, {\it {Neutrino oscillations and energy-momentum conservation}},
  {\em Mod. Phys. Lett. A} {\bf 25} (2010) 479--487,
  [\href{http://xxx.lanl.gov/abs/hep-ph/9604357}{{\tt hep-ph/9604357}}].

\bibitem{Nauenberg:1998vy}
M.~Nauenberg, {\it {Correlated wave packet treatment of neutrino and neutral
  meson oscillations}},  {\em Phys. Lett.} {\bf B447} (1999) 23--30,
  [\href{http://xxx.lanl.gov/abs/hep-ph/9812441}{{\tt hep-ph/9812441}}].

\bibitem{Cohen:2008qb}
A.~G. Cohen, S.~L. Glashow, and Z.~Ligeti, {\it {Disentangling Neutrino
  Oscillations}},  {\em Phys. Lett.} {\bf B678} (2009) 191--196,
  [\href{http://xxx.lanl.gov/abs/0810.4602}{{\tt arXiv:0810.4602}}].

\bibitem{Raghavan:2008cs}
R.~S. Raghavan, {\it {Hypersharp Resonant Capture of Anti-Neutrinos}},
  \href{http://xxx.lanl.gov/abs/0806.0839}{{\tt arXiv:0806.0839}}.

\bibitem{Litvinov:2008rk}
Y.~A. Litvinov {\em et.~al.}, {\it {Observation of Non-Exponential Orbital
  Electron Capture Decays of Hydrogen-Like $^{140}$Pr and $^{142}$Pm Ions}},
  {\em Phys. Lett.} {\bf B664} (2008) 162--168,
  [\href{http://xxx.lanl.gov/abs/0801.2079}{{\tt arXiv:0801.2079}}].

\bibitem{PhysRevA.76.052110}
W.~H. Zurek, {\it Quantum origin of quantum jumps: Breaking of unitary symmetry
  induced by information transfer in the transition from quantum to classical},
   {\em Phys. Rev. A} {\bf 76} (2007), no.~5 052110.

\bibitem{Eitel:1999gt}
K.~Eitel, {\it {Compatibility analysis of the LSND evidence and the KARMEN
  exclusion for $\bar\nu_\mu \to \bar\nu_e$ oscillations}},  {\em New J. Phys.}
  {\bf 2} (2000) 1, [\href{http://xxx.lanl.gov/abs/hep-ex/9909036}{{\tt
  hep-ex/9909036}}].

\bibitem{Gavrin:2010qj}
V.~Gavrin, V.~Gorbachev, E.~Veretenkin, and B.~Cleveland, {\it {Gallium
  experiments with artificial neutrino sources as a tool for investigation of
  transition to sterile states}},
  \href{http://xxx.lanl.gov/abs/1006.2103}{{\tt arXiv:1006.2103}}.

\bibitem{AguilarArevalo:2010wv}
{\bf The MiniBooNE} Collaboration, A.~A. Aguilar-Arevalo {\em et.~al.}, {\it
  {Event Excess in the MiniBooNE Search for $\bar \nu_\mu \rightarrow \bar
  \nu_e$ Oscillations}},  {\em Phys. Rev. Lett.} {\bf 105} (2010) 181801,
  [\href{http://xxx.lanl.gov/abs/1007.1150}{{\tt arXiv:1007.1150}}].

\bibitem{Athanassopoulos:1996jb}
{\bf LSND} Collaboration, C.~Athanassopoulos {\em et.~al.}, {\it {Evidence for
  $\bar{\nu}_\mu \to \bar{\nu}_e$ oscillations from the LSND experiment at the
  Los Alamos Meson Physics Facility}},  {\em Phys. Rev. Lett.} {\bf 77} (1996)
  3082--3085, [\href{http://xxx.lanl.gov/abs/nucl-ex/9605003}{{\tt
  nucl-ex/9605003}}].

\bibitem{Athanassopoulos:1997pv}
{\bf LSND} Collaboration, C.~Athanassopoulos {\em et.~al.}, {\it {Results on
  $\nu_\mu\to\nu_e$ neutrino oscillations from LSND}},  {\em Phys. Rev. Lett.}
  {\bf 81} (1998) 1774--1777,
  [\href{http://xxx.lanl.gov/abs/nucl-ex/9709006}{{\tt nucl-ex/9709006}}].

\bibitem{Giunti:2010zu}
C.~Giunti and M.~Laveder, {\it {Statistical Significance of the Gallium
  Anomaly}},  \href{http://xxx.lanl.gov/abs/1006.3244}{{\tt arXiv:1006.3244}}.

\bibitem{Mention:2011rk}
G.~Mention {\em et.~al.}, {\it {The Reactor Antineutrino Anomaly}},
  \href{http://xxx.lanl.gov/abs/1101.2755}{{\tt arXiv:1101.2755}}.

\bibitem{Yoshida:2006qz}
T.~Yoshida, T.~Kajino, H.~Yokomakura, K.~Kimura, A.~Takamura, {\em et.~al.},
  {\it {Supernova neutrino nucleosynthesis of light elements with neutrino
  oscillations}},  {\em Phys. Rev. Lett.} {\bf 96} (2006) 091101,
  [\href{http://xxx.lanl.gov/abs/astro-ph/0602195}{{\tt astro-ph/0602195}}].

\end{thebibliography}

\acknowledgments
One of us (DVA) thanks Hywel White for bringing him back to the subject of neutrino oscillations, and for the ensuing discussions. We thank Terry Goldman for reading a previous draft of this work, and Cheng-Yang Lee for discussions in the early stages of this communication. We also thank the Kaikoura and Westport Field Stations of the University of Canterbury, and Harish Chandra Research Institute, where much of this work was done.

\providecommand{\href}[2]{#2}\begingroup\raggedright\endgroup

\end{document}